\documentclass[%
 reprint,
 amsmath,amssymb,
 aps,
floatfix,
]{revtex4-1}

\usepackage{graphicx}
\usepackage{dcolumn}
\usepackage{bm}

\usepackage{qcircuit}

\usepackage{algorithm}
\usepackage[noend]{algpseudocode}
\algrenewcommand\algorithmicdo{}

\makeatletter
\renewcommand{\ALG@name}{Procedure}
\makeatother

\newcommand{\switch}{%
  \mathcode`+=\numexpr\mathcode`+ + "1000\relax 
  \mathcode`*=\numexpr\mathcode`* + "1000\relax
}

\usepackage[linktocpage=true,
  colorlinks=true, 
  pdfborder={0 0 0},
  linkcolor=blue,
  citecolor=red,
  filecolor=yellow,
  urlcolor=blue,
  bookmarks,
  pdfauthor={},
]{hyperref}

\begin{document}

\preprint{APS/123-QED}

\title{
Construction of Green's functions on a quantum computer:\\
Quasiparticle spectra of molecules
}

\author{Taichi Kosugi}

\author{Yu-ichiro Matsushita}
\affiliation{
Laboratory for Materials and Structures,
Institute of Innovative Research,
Tokyo Institute of Technology,
Yokohama 226-8503,
Japan
}

\date{\today}

\begin{abstract}
We propose a scheme for the construction of one-particle Green's function (GF) of an interacting electronic system via statistical sampling on a quantum computer.
Although the nonunitarity of creation and annihilation operators for the electronic spin orbitals prevents us from preparing specific states selectively,
probabilistic state preparation is demonstrated to be possible for the qubits.
We provide quantum circuits equipped with at most two ancillary qubits for obtaining all the components of GF.
We perform simulations of such construction of GFs for LiH and H$_2$O molecules based on the unitary coupled-cluster (UCC) method
to demonstrate the validity of our scheme by comparing the quasiparticle and satellite spectra exact within UCC and those from full configuration-interaction calculations.
We also examine the accuracy of sampling method by exploiting the Galitskii--Migdal formula, which gives the total energy only from the GF. 
\end{abstract}

\maketitle 

\section{Introduction}

Quantum chemistry calculations\cite{bib:4831} as a kind of quantum simulation\cite{bib:4828} have been drawing attention increasingly
since they serve as lucid proof-of-principle for quantum computation\cite{Nielsen_and_Chuang} and,
at the same time, are directly related to the state-of-the-art quantum hardware.
The electronic states and the operators acting on them for a given Hamiltonian have to be mapped to the qubits comprising a quantum computer and the operators for them by an appropriate transformation.\cite{bib:4294}
The Jordan--Wigner (JW)\cite{bib:4710} and Bravyi--Kitaev (BK)\cite{bib:4293} transformations are often used for quantum chemistry calculations.
Various approaches for obtaining the energy spectra of a many-electron system have been proposed.
The earliest one\cite{bib:4751} employs the quantum phase estimation (QPE) algorithm\cite{bib:4825, bib:4826} and the Suzuki--Trotter decomposition\cite{bib:4754} of the qubit Hamiltonian into a sequence of one- and two-qubit logic gates\cite{bib:4296} for unitary operations.
This approach was realized\cite{bib:4289} by using superconducting qubits.
Variational quantum eigensolver (VQE) is a newer approach,
in which a trial many-electron state is prepared via a quantum circuit with parameters to be optimized aiming at the ground state.
It uses a classical computer for updating the parameters based on the measurement results of the qubits,
which is why it is also called a quantum-classical hybrid algorithm.\cite{bib:4517}
This approach was first realized\cite{bib:4470} by using a quantum photonic device.
It has also been realized by superconducting\cite{bib:4289, bib:4292} and ion trap\cite{bib:4518} quantum computers.
Another approach for obtaining the energy spectra is the imaginary-time evolution.
It was recently proposed\cite{bib:4797, bib:4802, bib:4807} as a quantum-classical hybrid algorithm based on the McLachlan's variational principle.\cite{bib:4803}

An experiment of photoelectron spectroscopy (PES) irradiates light to a sample and
measures the energy of photoelectrons coming out of the sample.
An experiment of inverse PES is for the reverse process of PES.
Particularly for angle-resolved photoemission spectroscopy (ARPES),
the measured spectra of an interacting electronic system are often explained via the one-particle Green's function (GF).\cite{bib:4070,bib:4165,bib:pw_unfolding}
Since the GF contains rich information about the correlation effects in an electronic system\cite{bib:4575},
the GFs in the context of quantum chemistry\cite{Nooijen92,Nooijen93,Nooijen95,bib:3947,bib:4275} (on classical computers)
have been intensively studied recently for isolated\cite{bib:4483, bib:4516, bib:4582, bib:4829, bib:4830} and periodic\cite{bib:4473, bib:4723} systems. 
The reliable calculation of GFs is hence as important as that of the ground-state energies for molecular and solid-state systems.
It is, however, essentially expensive for classical computation
since it demands large memory and, often simultaneously,
large storage for description of an electronic state made up of lots of Slater determinants.
A quantum computer allows for, on the other hand,
representation of such an electronic state using the qubits thanks to the principle of superposition.
It is thus worth developing tools for electronic-structure calculations on quantum computers, which are coming to practical usage.

In this study,
we propose a scheme for the construction of one-particle GF of an interacting electronic system via statistical sampling on a quantum computer.
We introduce quantum circuits for probabilistic state preparation
which allow us to calculate the GF from the histogram obtained via measurements on the qubits.
Our scheme exploits the probabilistic preparation of electron-added and -removed states,
in contrast to the existing methods for GFs.\cite{bib:4611, bib:4617}
For demonstrating the validity of our scheme,
we perform simulations of such construction of molecular GFs based on the unitary coupled-cluster (UCC) method\cite{bib:4827}
by referring to the spectral functions exact within UCC.
We also examine the accuracy of sampling method by calculating the correlation energies from the GFs.

This paper is organized as follows.
In Section \ref{sec:methods},
we explain the theoretical perspective of our scheme.
In particular, we describe the quantum circuits in detail for obtaining GFs via statistical sampling.
In Section \ref{sec:computational_details},
we describe the computational details for our simulations on a classical computer.
In Section \ref{sec:results_and_discussion},
we show the simulation results of quantum computation for LiH and H$_2$O molecules.
In Section \ref{sec:conclusions},
we provide the conclusions.

\section{Methods}
\label{sec:methods}

We describe the scheme for constructing the GF using a quantum computer below in detail.
Although we use the ground states obtained in UCC calculations in the present study,
the scheme is applicable as long as the ground state can be prepared as the qubits.

\subsection{Unitary coupled-cluster method}

Let us consider an interacting $N$-electron system whose second-quantized Hamiltonian is $\mathcal{H}$.
A VQE calculation\cite{bib:4470} of quantum chemistry using the UCC method\cite{bib:4827} starts from an ansatz of the form
\begin{gather}
    U (\boldsymbol{\theta})
    =
        \exp [ T (\boldsymbol{\theta}) - T (\boldsymbol{\theta})^\dagger ]
        ,
    \label{def_UCC_opr}
\end{gather}
where $T (\boldsymbol{\theta})$ is an appropriately chosen cluster operator that depends on parameter(s) $\boldsymbol{\theta}$.
The transformation $U (\boldsymbol{\theta})$, which is unitary by definition,
is used to construct a trial ground state 
$
| \Psi (\boldsymbol{\theta}) \rangle
\equiv
U (\boldsymbol{\theta}) | \Psi_{\mathrm{ref}} \rangle
$
for given $\boldsymbol{\theta}$
from a reference state $| \Psi_{\mathrm{ref}} \rangle$.
In a practical VQE process,
the unitary transformation is implemented as parametrized operations on the qubits comprising a quantum computer.
The expected total energy
$E (\boldsymbol{\theta}) = \langle \Psi (\boldsymbol{\theta}) | \mathcal{H} | \Psi (\boldsymbol{\theta}) \rangle$
is obtained via measurements (Hamiltonian averaging\cite{bib:4289, bib:4518, bib:4470}), which is then used to update $\boldsymbol{\theta}$ iteratively according to an optimization scheme on a classical computer so that the measured energy at the next iteration is lower.
The optimized trial state will be used as the ground state $| \Psi_{\mathrm{gs}}^N \rangle$ for our scheme described below.

Although we introduce the electronic cluster operators for ans\"atze and rewrite them into the qubit representation in the present study,
one can instead start directly from ans\"atze given as qubit operators.
The qubit coupled-cluster\cite{bib:4713} is an approach in this direction.

\subsection{One-particle GFs}

\subsubsection{Definition}

Although we assume the ground state $| \Psi_{\mathrm{gs}}^N \rangle$ to be non-degenerate
and to be at zero temperature for simplicity,
the expressions provided below are easily extended for systems having degenerate ground states at nonzero temperature.
The one-particle GF\cite{fetter2003quantum, stefanucci2013nonequilibrium} of the system in frequency domain is given by
\begin{gather}
	G_{m m'} (z)
	=
		G_{m m'}^{(\mathrm{e}) } (z)
		+
		G_{m m'}^{(\mathrm{h}) } (z)
    \label{G_imag_freq_sum_partial_G_T0}
\end{gather}
for a complex frequency $z$,
where
\begin{gather}
	G_{m m'}^{(\mathrm{e}) } (z)
	=
		\langle \Psi_{\mathrm{gs}}^{N} | a_m 
		\frac{1}{ z + E_{\mathrm{gs}}^{N} - \mathcal{H}}
		a_{m'}^\dagger | \Psi_{\mathrm{gs}}^{N} \rangle
    \nonumber \\
	=
		\sum_{\lambda \in N + 1}
		\frac{
            B_{\lambda m m'}^{\mathrm{(e)}}
        }{
			z + E_\mathrm{gs}^N - E_\lambda^{N + 1}
		}
	\label{def_partial_G_e_T0}
\end{gather}
and
\begin{gather}
	G_{m m'}^{(\mathrm{h}) } (z)
	=
		\langle \Psi_{\mathrm{gs}}^N | a_{m'}^\dagger 
		\frac{1}{ z - E_{\mathrm{gs}}^N + \mathcal{H}}
		a_m | \Psi_{\mathrm{gs}}^N \rangle
    \nonumber \\
	=
		\sum_{\lambda \in N - 1}
		\frac{
            B_{\lambda m m'}^{\mathrm{(h)}}
        }{
			z + E_\lambda^{N - 1} - E_{\mathrm{gs}}^N
		}
	\label{def_partial_G_h_T0}
\end{gather}
are the electron- and hole-excitation parts of the GF, respectively.
$a^\dagger_m$ and $a_m$ are the creation and annihilation operators,
respectively,
of an electron at the $m$th spin orbital.
$E_{\mathrm{gs}}^N$ is the ground-state energy
and $E_\lambda^{N \pm 1}$ is the $\lambda$th energy eigenvalue of the $(N \pm 1)$-electron states.
\begin{gather}
    B_{\lambda m m'}^{\mathrm{(e)}}
    \equiv
        \langle \Psi_{\mathrm{gs}}^N  | a_m | \Psi_\lambda^{N + 1} \rangle
        \langle \Psi_\lambda^{N + 1} | a_{m'}^\dagger | \Psi_{\mathrm{gs}}^N \rangle
    \label{def_transition_mat_for_G_e}
\end{gather}
and
\begin{gather}
    B_{\lambda m m'}^{\mathrm{(h)}}
    \equiv
			\langle \Psi_{\mathrm{gs}}^N | a_{m'}^\dagger | \Psi_\lambda^{N - 1} \rangle
			\langle \Psi_\lambda^{N - 1} | a_m | \Psi_{\mathrm{gs}}^N \rangle
    \label{def_transition_mat_for_G_h}
\end{gather}
are the transition matrix elements.
The spectral function is defined via the GF as
\begin{gather}
    A (\omega)
    =
        -\frac{1}{\pi}
        \mathrm{Im Tr} \,
        G (\omega + i \delta)
    \label{def_spec_func}
\end{gather}
for a real $\omega$ with a small positive constant $\delta$ for ensuring causality.

It is clear from eqs. (\ref{def_partial_G_e_T0}) and (\ref{def_partial_G_h_T0}) that the calculation of GF requires not only the many-electron energy eigenvalues but also the transition matrix elements.
Various approaches for obtaining many-electron energy eigenvalues on a quantum computer have been proposed\cite{bib:4739, bib:4797, bib:4815, bib:4816, bib:4817}
and we can choose any alternative from them by comparing their precision and restriction from the viewpoints of algorithm and hardware.
As for the transition matrix elements, however,
there exists no established way for calculation of them on a quantum computer to our knowledge.
We therefore propose a scheme for the construction of GF via statistical sampling and describe it below in detail.
Our protocol is designed for obtaining the numerators on the RHSs in eqs. (\ref{def_partial_G_e_T0}) and (\ref{def_partial_G_h_T0}),
provided that the denominators have been known.

\subsubsection{Circuits for diagonal components}

In a typical scheme for the construction of GFs on a classical computer\cite{Nooijen92,Nooijen93,Nooijen95,bib:3947,bib:4275},
the equation-of-motion coupled-cluster (EOM-CC) approach is adopted to obtain the energy eigenvalues and the transition matrix elements for the $(N \pm 1)$-electron intermediate states.
In the present case, one might think by looking at eq. (\ref{def_transition_mat_for_G_e}) that
$B_{\lambda m m'}^{\mathrm{(e)}}$ can be easily calculated by preparing the qubit representations of $| \Psi_\lambda^{N + 1} \rangle$ and
$
| \Psi_m^{\mathrm{(e)}} \rangle
\equiv
a_m^\dagger | \Psi_{\mathrm{gs}}^N \rangle
,
$
between which the inner product is calculated using the \textsc{swap} test or its versions\cite{bib:4738, bib:4739, bib:4740} with phase factors.
Such an approach is, however, difficult in fact.
It is because the creation operator is not unitary and the norm of the electron-added state is not conserved in general, that is,
$\langle \Psi_m^{\mathrm{(e)}} | \Psi_m^{\mathrm{(e)}} \rangle \ne 1$.
This fact prevents one from preparing a specific electron-added state selectively
since a quantum circuit can apply only unitary operations to qubits.
This difficulty is similarly the case for the electron-removed (hole-added) state
$
| \Psi_m^{\mathrm{(h)}} \rangle
\equiv
a_m | \Psi_{\mathrm{gs}}^N \rangle
$
.
To circumvent this difficulty,
we have to resort to another approach.

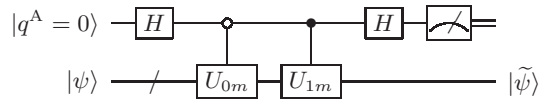
\begin{figure}
\centering
\mbox{ 
\Qcircuit @C=1em @R=1em { 
    \lstick{| q^{\mathrm{A}} = 0 \rangle} & \gate{H} & \ctrlo{1}  & \ctrl{1}   & \gate{H} & \meter & \cw \\
    \lstick{| \psi \rangle}      & {/} \qw  & \gate{U_{0 m}} & \gate{U_{1 m}} & \qw      & \qw    & \rstick{| \widetilde{\psi} \rangle} \qw
} 
} 
\vspace{0.5cm}
\caption{
Diagonal circuit $\mathcal{C}_m$ for probabilistic preparation of
$a_m^\dagger | \psi \rangle$ 
and
$a_m | \psi \rangle$
from an arbitrary input state $| \psi \rangle$ and an ancillary qubit $|q^{\mathrm{A}} \rangle$ using the unitary operations $U_{0 m}$ and $U_{1 m}$.
$H$ in the circuit represents the Hadamard gate.
} 
\label{circuit_prep_non_unitary} 
\end{figure}

As explained in Introduction,
the JW\cite{bib:4710} and BK\cite{bib:4293} transformations are often used for mapping a many-electron state to a many-qubit state.
We do not distinguish between the kets as many-electron states and those as many-qubit states in what follows since such simplification will not cause confusion for the readers.
By looking at the definitions of the transformations
[see, e.g., eqs. (34), (39), and (40) in Ref.\cite{bib:4294}],
we can notice that for both transformations any pair of electronic creation and  annihilation operators can be expressed
by using two unitary operators $U_{0 m}$ and $U_{1 m}$ on qubits as
\begin{gather}
    a_m^\dagger
    =
        \frac{U_{0 m} - U_{1 m}}{2}
    \label{el_oprs_using_unitary_c}
\end{gather}
and
\begin{gather}
    a_m
    =
        \frac{U_{0 m} + U_{1 m}}{2}
    \label{el_oprs_using_unitary_a}
\end{gather}
for a given $m$,
regardless of the number of qubits comprising the quantum computer.
Such a decomposition of electronic operators is in fact always possible
since $a_m^\dagger + a_m$ and $a_m^\dagger - a_m$ are ensured to be unitary
by the anti-commutation relation,
reminding us of the Majorana fermions.\cite{bib:4979}
We introduce a trick for state preparation by exploiting this fact.
Specifically, we construct a circuit $\mathcal{C}_m$ equipped with an ancillary qubit $| q^{\mathrm{A}} \rangle$ by implementing the controlled operations of $U_{0 m}$ and $U_{1 m}$,
as shown in Figure \ref{circuit_prep_non_unitary}. 
The whole system consists of the ancilla and an arbitrary input register $| \psi \rangle$,
whose state changes by undergoing the circuit as
\begin{gather}
    | 0 \rangle 
    \otimes
    | \psi \rangle
    \longmapsto
        | 0 \rangle
        \otimes
        \frac{U_{0 m} + U_{1 m}}{2}
        | \psi \rangle
        +
        | 1 \rangle
        \otimes
        \frac{U_{0 m} - U_{1 m}}{2}
        | \psi \rangle
    \nonumber \\
    =
        | 0 \rangle
        \otimes
        a_m
        | \psi \rangle
        +
        | 1 \rangle
        \otimes
        a_m^\dagger
        | \psi \rangle
    \equiv
        | \Phi_m \rangle
    .
    \label{GF_circuit_e_and_h_diag}
\end{gather}
The action of the circuit to the whole system is easily confirmed to be unitary due to the anti-commutation relation between the electronic operators.
The projective measurement\cite{Nielsen_and_Chuang} on the ancillary bit is represented by the two operators
$\mathcal{P}_q = | q \rangle \langle q | \otimes I \, (q = 0, 1)$,
for which $| q \rangle $ is observed with a probability $\langle \Phi_m | \mathcal{P}_q | \Phi_m \rangle$.
The state of the whole system collapses immediately after the measurement as follows:
\begin{gather}
    | \Phi_m \rangle
    \overset{| 0 \rangle \, \mathrm{observed}}{\longmapsto}
        | 0 \rangle
        \otimes
        \frac{a_m}{ \sqrt{p_m (\mathrm{h})} }
        | \psi \rangle
    \nonumber \\
    \mathrm{prob.}
    \,
        \langle \psi | a_m^\dagger a_m | \psi \rangle
    \equiv
        p_m (\mathrm{h})
    \label{GF_state_transition_h}
    \\
    | \Phi_m \rangle
    \overset{| 1 \rangle \, \mathrm{observed}}{\longmapsto}
        | 1 \rangle
        \otimes
        \frac{a_m^\dagger}{ \sqrt{p_m (\mathrm{e})}}
        | \psi \rangle
    \nonumber \\
    \mathrm{prob.}
    \,
        \langle \psi | a_m a_m^\dagger | \psi \rangle
    \equiv
        p_m (\mathrm{e})
    \label{GF_state_transition_e}
\end{gather}
This result implies that $\mathcal{C}_m$ allows us to prepare the two states $a_m^\dagger | \psi \rangle$ and $a_m | \psi \rangle$ probabilistically apart from their normalization constants.
For the number $N_{\mathrm{meas}}$ of measurements on the ancilla,
the probability distribution of counted outcomes for $| 0 \rangle$, or equivalently $| 1 \rangle$,
is a binomial distribution.
The probability distribution thus converges to the normal distribution for many repeated measurements
and the error of normalization constant scales as $N_{\mathrm{meas}}^{-1/2}$.
This circuit is used for obtaining the diagonal components of the GF,
as explained later.

\subsubsection{Circuits for off-diagonal components}

For the $m$th and $m'$th spin orbitals ($m \ne m'$),
we define the following auxiliary creation and annihilation operators
\begin{gather}
    a_{m m'}^{\pm}
    \equiv
        \frac{
            a_m
            \pm
            e^{-i \pi/4}
            a_{m'}
        }{2}
    \label{GF_def_annihilation_aux}
\end{gather}
and
\begin{gather}
    a_{m m'}^{\pm \dagger}
    \equiv
        \frac{
            a_m^\dagger
            \pm
            e^{i \pi/4}
            a_{m'}^\dagger
        }{2}
        ,
    \label{GF_def_creation_aux}
\end{gather}
respectively,
which are the Hermitian conjugate of each other.
Unnormalized auxiliary $(N + 1)$-electron states
\begin{gather}
    | \Psi_{m m'}^{\mathrm{(e)} \pm} \rangle
    \equiv
        a_{m m'}^{\pm \dagger}
        | \Psi_{\mathrm{gs}}^N \rangle
    \label{def_aux_state_e}
\end{gather}
can have overlaps with the energy eigenstates as
\begin{gather}
    D_{\lambda m m'}^{\mathrm{(e)} \pm}
    \equiv
        | \langle \Psi_\lambda^{N + 1} | \Psi_{m m'}^{\mathrm{(e)} \pm} \rangle |^2
    \nonumber \\
    =
        \frac{
            B_{\lambda m m}^{\mathrm{(e)}}
            +
            B_{\lambda m' m'}^{\mathrm{(e)}}
            }{4}
        \pm
        \frac{
            e^{i \pi/4} B_{\lambda m m'}^{\mathrm{(e)}}
            +
            e^{-i \pi/4} B_{\lambda m' m}^{\mathrm{(e)}}
        }{4}
        .
    \label{def_transition_mat_for_G_e_aux}
\end{gather}
By solving eq. (\ref{def_transition_mat_for_G_e_aux}) for the off-diagonal component of $B_{\lambda}^{\mathrm{(e)}}$,
we can calculate it from $D_{\lambda}^{\mathrm{(e)} \pm}$ and the diagonal components of $B_{\lambda}^{\mathrm{(e)}}$ as
\begin{gather}
    B_{\lambda m m'}^{\mathrm{(e)}}
    =
        \pm
        \Bigg(
            2
            e^{-i \pi/4}
            D_{\lambda m m'}^{\mathrm{(e)} \pm}
            +
            2
            e^{i \pi/4}
            D_{\lambda m' m}^{\mathrm{(e)} \pm}
    \nonumber \\
            -
            \frac{
                B_{\lambda m m}^{\mathrm{(e)}}
                +
                B_{\lambda m' m'}^{\mathrm{(e)}}
            }{\sqrt{2}}
        \Bigg)
        .
    \label{transition_G_e_off_diag}
\end{gather}
From the two expressions for both signs in eq. (\ref{transition_G_e_off_diag}),
we can obtain the off-diagonal component of $B_{\lambda}^{\mathrm{(e)}}$ only from $D_{\lambda}^{\mathrm{(e)} \pm}$ as
\begin{gather}
    B_{\lambda m m'}^{\mathrm{(e)}}
    =
        e^{-i \pi/4}
        (
            D_{\lambda m m'}^{\mathrm{(e)} +}
            -
            D_{\lambda m m'}^{\mathrm{(e)} -}
        )
    \nonumber \\
        +
        e^{i \pi/4}
        (
            D_{\lambda m' m}^{\mathrm{(e)} +}
            -
            D_{\lambda m' m}^{\mathrm{(e)} -}
        )
        .
    \label{transition_G_e_off_diag_alpha_half}
\end{gather}

For unnormalized auxiliary $(N - 1)$-electron states
\begin{gather}
    | \Psi_{m m'}^{\mathrm{(h)} \pm} \rangle
    \equiv
        a_{m m'}^{\pm}
        | \Psi_{\mathrm{gs}}^N \rangle
        ,
    \label{def_aux_state_h}
\end{gather}
the expression of
$
D_{\lambda m m'}^{\mathrm{(h)} \pm}
\equiv
| \langle \Psi_\lambda^{N - 1} | \Psi_{m m'}^{\mathrm{(h)} \pm} \rangle |^2
$
is the same as that in eq. (\ref{def_transition_mat_for_G_e_aux}) with $(\mathrm{e})$ replaced by $(\mathrm{h})$.
This means that we can calculate the off-diagonal component of $B_{\lambda}^{\mathrm{(h)}}$ from $D_{\lambda}^{\mathrm{(h)} \pm}$ by using the same expression as eq. (\ref{transition_G_e_off_diag_alpha_half}) with the replacement.

\begin{figure*}
\centering
\mbox{ 
\Qcircuit @C=0.5em @R=1em { 
    \lstick{| q^{\mathrm{A}}_0 = 0 \rangle} & \gate{H} & \ctrlo{1}  & \ctrl{1}    & \qw                 & \ctrlo{1}  & \ctrl{1}   & \gate{H} & \meter & \cw \\
    \lstick{                    | q^{\mathrm{A}}_1 = 0 \rangle} & \gate{H} & \ctrlo{1}  & \ctrlo{1}   & \gate{Z(\pi/4)} & \ctrl{1}   & \ctrl{1}   & \gate{H} & \meter & \cw \\
    \lstick{| \psi \rangle}        & {/} \qw  & \gate{U_{0 m}} & \gate{U_{1 m}}  & \qw                 & \gate{U_{0 m'}} & \gate{U_{1 m'}} & \qw      & \qw    & \rstick{| \widetilde{\psi} \rangle} \qw
} 
} 
\vspace{0.5cm}
\caption{
Off-diagonal circuit $\mathcal{C}_{m m'}$ for probabilistic preparation of
$a_{m m'}^{\pm \dagger} | \psi \rangle$
and
$a_{m' m}^{\pm} | \psi \rangle$
from an arbitrary input state $| \psi \rangle$ and two ancillary qubits $| q^{\mathrm{A}}_0 \rangle$ and $| q^{\mathrm{A}}_1 \rangle$ using the unitary operations $U_{0 m}, U_{1 m}, U_{0 m'}$, and $U_{1 m'}$.
$Z (\pi/4) = \mathrm{diag}(1, e^{i \pi/4})$ is a phase gate.
} 
\label{circuit_prep_non_unitary_aux} 
\end{figure*}
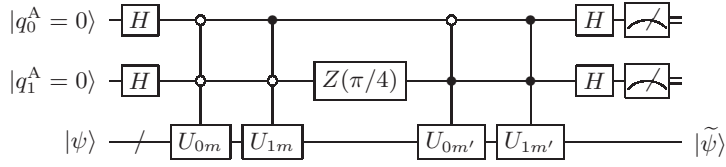

We construct a circuit $\mathcal{C}_{m m'}$ equipped with two ancillary qubits $| q_0^{\mathrm{A}} \rangle$ and $| q_1^{\mathrm{A}} \rangle$ by implementing the controlled operations of $U_{0 m}, U_{1 m}, U_{0 m'}$, and $U_{1 m'}$,
as shown in Figure \ref{circuit_prep_non_unitary_aux}. 
The whole system consists of the ancillae and an arbitrary input register $| \psi \rangle$,
whose state changes by undergoing the circuit as
\begin{gather}
        | q^{\mathrm{A}}_1 = 0 \rangle
        \otimes
        | q^{\mathrm{A}}_0 = 0 \rangle
    \otimes
        | \psi \rangle
    \nonumber \\
    \longmapsto
            | 0 \rangle \otimes | 0 \rangle \otimes
                \frac{U_{0 m} + U_{1 m}
                        +
                        e^{i \pi/4} (U_{0 m'} + U_{1 m'})}{4}
            | \psi \rangle
    \nonumber \\
            +
            | 0 \rangle \otimes | 1 \rangle \otimes
                \frac{U_{0 m} - U_{1 m}
                        +
                        e^{i \pi/4} (U_{0 m'} - U_{1 m'})}{4}
            | \psi \rangle
    \nonumber \\
            +
            | 1 \rangle \otimes | 0 \rangle \otimes
                \frac{U_{0 m} + U_{1 m}
                        -
                        e^{i \pi/4} (U_{0 m'} + U_{1 m'}) }{4}
            | \psi \rangle
    \nonumber \\
            +
            | 1 \rangle \otimes | 1 \rangle \otimes
                \frac{ U_{0 m} - U_{1 m}
                        -
                        e^{i \pi/4} (U_{0 m'} - U_{1 m'}) }{4}
            | \psi \rangle
    \nonumber \\
    =
            | 0 \rangle \otimes | 0 \rangle \otimes
            e^{i \pi/4} a_{m' m}^+
            | \psi \rangle
            +
            | 0 \rangle \otimes | 1 \rangle \otimes
            a_{m m'}^{+ \dagger} 
            | \psi \rangle
    \nonumber \\
            -
            | 1 \rangle \otimes | 0 \rangle \otimes
            e^{i \pi/4} a_{m' m}^- 
            | \psi \rangle
            +
            | 1 \rangle \otimes | 1 \rangle \otimes
            a_{m m'}^{- \dagger}
            | \psi \rangle
    \nonumber \\
    \equiv
        | \Phi_{m m'} \rangle
        .
\end{gather}
The action of the circuit to the whole system is easily confirmed to be unitary due to the anti-commutation relation between the electronic operators.
The projective measurement on the ancillary bits is represented by the four operators
$\mathcal{P}_{q q'} = | q \rangle \langle q | \otimes | q' \rangle \langle q' | \otimes I \, (q, q' = 0, 1)$,
for which $| q \rangle \otimes | q' \rangle$ is observed with a probability $\langle \Phi_{m m'} | \mathcal{P}_{q q'} | \Phi_{m m'} \rangle$.
The state of the whole system collapses immediately after the measurement as follows:
\begin{gather}
    | \Phi_{m m'} \rangle
    \overset{| 0 \rangle \otimes | 0 \rangle \, \mathrm{observed}}{\longmapsto}
        | 0 \rangle
        \otimes
        | 0 \rangle
        \otimes
        \frac{a_{m' m}^+}{\sqrt{p_{m m'} (\mathrm{h}, +)}}
        | \psi \rangle
    \nonumber \\
    \mathrm{prob.}
    \,
        \langle \psi |
            a_{m' m}^{+ \dagger}
            a_{m' m}^+
        | \psi \rangle
    \equiv
        p_{m m'} (\mathrm{h}, +)
    \label{GF_state_transition_h_p}
    \\
    | \Phi_{m m'} \rangle
    \overset{| 0 \rangle \otimes | 1 \rangle \, \mathrm{observed}}{\longmapsto}
        | 0 \rangle
        \otimes
        | 1 \rangle
        \otimes
        \frac{a_{m m'}^{+ \dagger}}{\sqrt{p_{m m'} (\mathrm{e}, +)}}
        | \psi \rangle
    \nonumber \\
    \mathrm{prob.}
    \,
        \langle \psi |
            a_{m m'}^+
            a_{m m'}^{+ \dagger}
        | \psi \rangle
    \equiv
        p_{m m'} (\mathrm{e}, +)
    \label{GF_state_transition_e_p}
    \\
    | \Phi_{m m'} \rangle
    \overset{| 1 \rangle \otimes | 0 \rangle \, \mathrm{observed}}{\longmapsto}
        | 1 \rangle
        \otimes
        | 0 \rangle
        \otimes
        \frac{a_{m' m}^- }{\sqrt{p_{m m'} (\mathrm{h}, -)}}
        | \psi \rangle
    \nonumber \\
    \mathrm{prob.}
    \,
        \langle \psi |
            a_{m' m}^{- \dagger}
            a_{m' m}^-
        | \psi \rangle
    \equiv
        p_{m m'} (\mathrm{h}, -)
    \label{GF_state_transition_h_m}
    \\
    | \Phi_{m m'} \rangle
    \overset{| 1 \rangle \otimes | 1 \rangle \, \mathrm{observed}}{\longmapsto}
        | 1 \rangle
        \otimes
        | 1 \rangle
        \otimes
        \frac{a_{m m'}^{- \dagger} }{\sqrt{p_{m m'} (\mathrm{e}, -)}}
        | \psi \rangle
    \nonumber \\
    \mathrm{prob.}
    \,
        \langle \psi |
            a_{m m'}^-
            a_{m m'}^{- \dagger}
        | \psi \rangle
    \equiv
        p_{m m'} (\mathrm{e}, -)
    \label{GF_state_transition_e_m}
\end{gather}
This result implies that $\mathcal{C}_{m m'}$ allows us to prepare the four states
$a_{m m'}^{\pm \dagger} | \psi \rangle$
and
$a_{m' m}^\pm | \psi \rangle$
probabilistically apart from their normalization constants.
For the number $N_{\mathrm{meas}}$ of measurements on the ancillae,
the error of normalization constant scales as $N_{\mathrm{meas}}^{-1/2}$
similarly to the case for diagonal components.
This circuit is used for obtaining the off-diagonal components of the GF,
as explained below.

\subsubsection{Transition matrices via statistical sampling}

Given the results of measurement on the ancillary bit(s),
we have the register $| \widetilde{\psi} \rangle$ representing the $N_e$-electron state with $N_e = N + 1$ or $N - 1$.
Then we perform QPE by inputting $| \widetilde{\psi} \rangle$ to obtain the energy eigenvalue in the subspace spanned by the $N_e$-electron states.
A QPE experiment inevitably suffers from probabilistic errors that depend on the number and the initial states of qubits.\cite{Nielsen_and_Chuang, bib:4716}
Furthermore,
the results are affected by the number of steps for the Suzuki--Trotter decomposition and the order of partial Hamiltonians.
We assume for simplicity, however, that the QPE procedure is realized on a quantum computer with ideal precision.
We will thus find the estimated value to be $E_{\lambda}^{N_e}$ with a probability
$| \langle \Psi_\lambda^{N_e} | \widetilde{\psi} \rangle |^2$.
\cite{Nielsen_and_Chuang}

If we input $| \Psi_{\mathrm{gs}}^N \rangle$ to the diagonal circuit $\mathcal{C}_m$ in Fig. \ref{circuit_prep_non_unitary} and
process the whole system the way described above,
the energy eigenvalue $E_{\lambda}^{N + 1}$ will be obtained with a probability
[see eq. (\ref{GF_state_transition_e})]
\begin{gather}
    p_m (E_{\lambda}^{N + 1})
    =
        \left| \langle \Psi_\lambda^{N + 1} | \frac{a_m^\dagger}{\sqrt{p_m (\mathrm{e})}} | \Psi_{\mathrm{gs}}^N \rangle \right|^2
        p_m (\mathrm{e})
    =
        B_{\lambda m m}^{\mathrm{(e)}}
    ,
    \label{prob_energy_diag_el}
\end{gather}
while $E_{\lambda}^{N - 1}$ will be obtained with a probability
[see eq. (\ref{GF_state_transition_h})]
\begin{gather}
    p_m (E_{\lambda}^{N - 1})
    =
        \left| \langle \Psi_\lambda^{N - 1} | \frac{a_m}{\sqrt{p_m (\mathrm{h})}} | \Psi_{\mathrm{gs}}^N \rangle \right|^2
        p_m (\mathrm{h})
    =
        B_{\lambda m m}^{\mathrm{(h)}}
        .
\end{gather}
This means that we can get the diagonal components of transition matrices $B_\lambda^{(\mathrm{e})}$ and $B_\lambda^{(\mathrm{h})}$ via statistical sampling for a fixed $m$.
It is easily confirmed that 
$
\sum_{\lambda \in N - 1} p_m (E_{\lambda}^{N - 1})
+
\sum_{\lambda \in N + 1} p_m (E_{\lambda}^{N + 1})
=
1
$
due to the completeness of
$\{ | \Psi_\lambda^{N + 1} \rangle \}_\lambda$
for the $(N + 1)$-electron states and that of
$\{ | \Psi_\lambda^{N - 1} \rangle \}_\lambda$
for the $(N - 1)$-electron states, as expected.

If we input $| \Psi_{\mathrm{gs}}^N \rangle$ to the off-diagonal circuit $\mathcal{C}_{m m'}$ in Fig. \ref{circuit_prep_non_unitary_aux} and
process the whole system the way described above,
the ancillary bits $| 0 \rangle \otimes | 1 \rangle$ or $| 1 \rangle \otimes | 1 \rangle$ will be observed and
the energy eigenvalue $E_{\lambda}^{N + 1}$ will be obtained with probabilities
[see eqs. (\ref{GF_state_transition_e_p}) and (\ref{GF_state_transition_e_m})]
\begin{gather}
    p_{m m'} (\pm, E_{\lambda}^{N + 1})
    \nonumber \\
    =
        \left| \langle \Psi_\lambda^{N + 1} |
            \frac{a_{m m'}^{\pm \dagger} }
                {\sqrt{p_{m m'} (\mathrm{e}, \pm)}}
            | \Psi_{\mathrm{gs}}^N \rangle \right|^2
        p_{m m'} (\mathrm{e}, \pm)
    =
        D_{\lambda m m'}^{\mathrm{(e)} \pm}
        ,
\end{gather}
while the ancillary bits $| 0 \rangle \otimes | 0 \rangle$ or $| 1 \rangle \otimes | 0 \rangle$ will be observed and
the energy eigenvalue $E_{\lambda}^{N - 1}$ will be obtained with probabilities
[see eqs. (\ref{GF_state_transition_h_p}) and (\ref{GF_state_transition_h_m})]
\begin{gather}
    p_{m m'} (\pm, E_{\lambda}^{N - 1})
    \nonumber \\
    =
        \left| \langle \Psi_\lambda^{N - 1} |
            \frac{a_{m' m}^{\pm} }
                {\sqrt{p_{m m'} (\mathrm{h}, \pm)}}
            | \Psi_{\mathrm{gs}}^N \rangle \right|^2
        p_{m m'} (\mathrm{h}, \pm)
    =
        D_{\lambda m m'}^{\mathrm{(h)} \pm}
        .
    \label{prob_energy_off_diag_hole}
\end{gather}
This means that we can get the off-diagonal components of transition matrices $B_\lambda^{(\mathrm{e})}$ and $B_\lambda^{(\mathrm{h})}$
from eq. (\ref{transition_G_e_off_diag_alpha_half})
via statistical sampling for a fixed combination of $m$ and $m'$.
It is easily confirmed that 
$
\sum_{\sigma = +, -}
[
\sum_{\lambda \in N - 1}
p_{m m'} (\sigma, E_{\lambda}^{N - 1})
+
\sum_{\lambda \in N + 1}
p_{m m'} (\sigma, E_{\lambda}^{N + 1})
]
=
1
,
$
as expected.

We provide the pseudocodes in Appendix \ref{appendix:pseudocodes} for the calculation process of GF explained above.

\subsection{Galitskii--Migdal formula}

The Galitskii--Migdal (GM) formula\cite{fetter2003quantum} enables one to calculate the ground-state energy of an interacting electronic system solely from the time-ordered GF.
It can be rewritten to a tractable form for
representation using restricted Hartree--Fock (RHF) orbitals as\cite{bib:3575, bib:3576, bib:3581}
\begin{gather}
    E_{\mathrm{GM}} [G]
    =
        E_{\mathrm{nucl}}
        +
        \frac{1}{2}
        \sum_\sigma
        \mathrm{Tr}
            [
            (h + \varepsilon) 
            \gamma_\sigma
            ]
    \nonumber \\
        +
        \frac{1}{2}
        \sum_\sigma
        \frac{1}{2 \pi i}
        \int_{-\infty}^\infty
        d \omega \,
            e^{+i \omega 0}
            \mathrm{Tr}
            [
                \Sigma_{\mathrm{c} \sigma} (\omega)
                G_\sigma (\omega)
            ]
        ,
    \label{GM_formula_for_HF}
\end{gather}
where the integrand on the RHS contains a convergence factor $e^{+i \omega 0}$,
forcing us to pick up the poles of the GF for the states below the Fermi level.
$E_{\mathrm{nucl}}$ is the nuclear-repulsion energy.
For spatial HF orbitals $p$ and $p'$,
$h_{p p'}$ is the matrix element of
the one-electron operator $h (\boldsymbol{r})$,
which is the sum of the kinetic-energy term and the ionic-potential term.
$\varepsilon$ is the diagonal matrix whose components are the HF orbital energies.
\begin{gather}
    \gamma_{\sigma p p'}
    \equiv
        \langle a_{\sigma p'}^\dagger a_{\sigma p} \rangle
    =
        \frac{1}{2 \pi i}
        \int_{-\infty}^\infty
        d \omega \,
            e^{+i \omega 0}
            G_{\sigma p p'} (\omega)
    \label{def_dens_mat}
\end{gather}
is the one-particle density matrix\cite{fetter2003quantum,stefanucci2013nonequilibrium} for spin $\sigma$.
$\Sigma_{\mathrm{c}}$ is the self-energy obtained from the Dyson equation
$
\Sigma_{\mathrm{c} \sigma} [G]
=
G_{\mathrm{HF} \sigma}^{-1} 
-
G_{\sigma}^{-1}
,
$
where the HF GF is given solely by the orbital energies:
$
G_{\mathrm{HF} \sigma p p'} (\omega)
=
\delta_{p p'}
(\omega - \varepsilon_p)^{-1}
.
$
$\Sigma_{\mathrm{c}}$ is responsible for the correlation effects in the (interacting) GF,
which are not taken into account in the HF solution.
We can use the expression for $E_{\mathrm{GM}}$ as an energy functional for an arbitrary input GF.
If we substitute the HF GF into eq. (\ref{GM_formula_for_HF}),
the third term on the RHS vanishes and
we get the well known expression for the HF total energy,
$
E_{\mathrm{HF}}
=
E_{\mathrm{GM}} [G_{\mathrm{HF}}]
=
E_{\mathrm{nucl}}
+
\sum_\sigma
\mathrm{Tr}
[
(h + \varepsilon) 
\gamma_{\mathrm{HF} \sigma}
]/2
,
$
where $\gamma_{\mathrm{HF} \sigma}$ is the HF density matrix.
The total energy for the interacting case is thus written as
$
E_{\mathrm{GM}} [G]
=
E_{\mathrm{HF}}
+
\Delta E_1 [G]
+
\Delta E_2 [G]
,
$
where the sum of
\begin{gather}
    \Delta E_1 [G]
    \equiv
        \frac{1}{2}
        \sum_\sigma
        \mathrm{Tr}
            [
            (h + \varepsilon) 
            ( \gamma_\sigma - \gamma_{\mathrm{HF} \sigma})
            ]
    \label{GM_formula_Delta_E_1}
\end{gather}
and
\begin{gather}
    \Delta E_2 [G]
    \equiv
        \frac{1}{2}
        \sum_\sigma
        \frac{1}{2 \pi i}
        \int_{-\infty}^\infty
        d \omega \,
            e^{+i \omega 0}
            \mathrm{Tr}
            [
                \Sigma_{\mathrm{c} \sigma} (\omega)
                G_\sigma (\omega)
            ]
        .
    \label{GM_formula_Delta_E_2}
\end{gather}
is the correlation energy.
$\Delta E_1$ is interpreted as the energy correction coming from the variation in the occupancy of HF orbitals,
while an interpretation for $\Delta E_2$ within the HF picture is difficult to draw.
We should keep in mind that
$E_{\mathrm{GM}} [G_{\mathrm{trial}}]$ calculated from the GF $G_{\mathrm{trial}}$ for a trial ground state $| \Psi_{\mathrm{trial}} \rangle$
via eqs. (\ref{def_partial_G_e_T0}) and (\ref{def_partial_G_h_T0}) can differ from the expected energy in general:
$
E_{\mathrm{GM}} [G_{\mathrm{trial}}]
\ne
\langle \Psi_{\mathrm{trial}} | \mathcal{H} | \Psi_{\mathrm{trial}} \rangle
$
,
since eqs. (\ref{def_partial_G_e_T0}) and (\ref{def_partial_G_h_T0}) use the fact that
the true ground state is an eigenstate of $\mathcal{H}$.
We use the expressions in eqs. (\ref{GM_formula_Delta_E_1}) and (\ref{GM_formula_Delta_E_2}), however,
to examine quantitatively the accuracy of GFs calculated in the present study.
It is because one of our purposes is to see how $E_{\mathrm{GM}}$ values for UCC GFs from statistical sampling approach the ideal values as the number of measurements increases.

\section{Computational details}
\label{sec:computational_details}

We adopted STO-3G basis sets as the Cartesian Gaussian-type basis functions\cite{Helgaker} for all the elements in our quantum chemistry calculations.
The Coulomb integrals between the atomic orbitals were calculated efficiently.\cite{Libint1}
We first performed RHF calculations to get the orthonormalized molecular orbitals in the target systems and
calculated the two-electron integrals between them,
from which we constructed the second-quantized electronic Hamiltonians.
After that,
we used JW transformation to get the Hamiltonians in qubit representation by using OpenFermion\cite{OpenFermion} to perform full configuration-interaction (FCI) and UCC calculations.
The parameters in the UCC calculations were optimized by employing the constrained optimization by linear approximation (COBYLA) method.

Although our scheme for the calculation of GF assumes that the energy spectra of $(N \pm 1)$-electron states for a target system are already known (see Procedure \ref{alg:CalcGF}),
which can be obtained in various approaches for quantum computers,\cite{bib:4739, bib:4797, bib:4815, bib:4816, bib:4817}
we simply use those obtained in (classical) FCI calculations for the $(N \pm 1)$-electron states in the present study.
It is because the main purpose is to demonstrate succinctly the validity of our scheme for GFs using statistical sampling.
Simulations of GFs by taking into account the restrictions on the accuracy of spectra of excited states imposed by hardware and/or specific algorithms should be performed in the future.
Our calculations of GFs, including those simulated with statistical sampling,
were performed by substituting the necessary quantities into the Lehmann representation,
given by eqs. (\ref{def_partial_G_e_T0}) and (\ref{def_partial_G_h_T0}).
We set $\delta$ in eq. (\ref{def_spec_func}) to $0.02$ a.u. for the spectral functions throughout the present study.

For numerical evaluation of the integrals in eqs. (\ref{def_dens_mat}) and (\ref{GM_formula_Delta_E_2}),
we adopted rectangular contours on the complex plane
so that they encircle all the poles on the negative real axis for the integrands.

\section{Results and discussion}
\label{sec:results_and_discussion}

\subsection{LiH molecule}

\subsubsection{UCC calculations}

\begin{figure}
\begin{center}
\includegraphics[width=5.5cm]{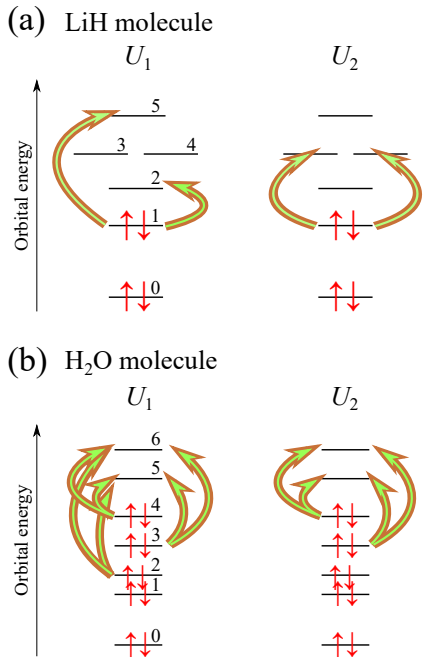}
\end{center}
\caption{
Schematic illustration of RHF orbitals and their electronic occupancies for (a) an LiH molecule and (b) an H$_2$O molecule.
Integers near the individual orbitals are the orbital indices.
The orbital $0$ is contributed mainly from the $1 s$ orbital of Li in (a) and O in (b).
Thick arrows represent the excitation channels in $T_1$ and $T_2$ for the ans\"atze $U_1$ and $U_2$, respectively, introduced in the UCC calculations. 
}
\label{fig:orb_energies}
\end{figure}

\begin{figure*}
\centering
\mbox{ 
\Qcircuit @C=0.5em @R=0.5em { 
    \lstick{| q_0 = 0 \rangle}    & \gate{X} & \qw      & \qw      & \qw      & \qw      & \qw                   & \qw      & \qw      & \qw      & \qw              & \qw      & \qw      & \qw      & \qw      & \qw                   & \qw      & \qw      & \qw      & \qw              & \qw \\
    \lstick{| q_1 = 0 \rangle}    & \gate{X} & \qw      & \qw      & \qw      & \qw      & \qw                   & \qw      & \qw      & \qw      & \qw              & \qw      & \qw      & \qw      & \qw      & \qw                   & \qw      & \qw      & \qw      & \qw              & \qw \\
    \lstick{| q_2 = 0 \rangle}    & \gate{X} & \gate{H} & \ctrl{1} & \qw      & \qw      & \qw                   & \qw      & \qw      & \ctrl{1} & \gate{H}         & \gate{H} & \ctrl{1} & \qw      & \qw      & \qw                   & \qw      & \qw      & \ctrl{1} & \gate{H}         & \qw \\
    \lstick{| q_3 = 0 \rangle}    & \gate{X} & \gate{H} & \targ    & \ctrl{1} & \qw      & \qw                   & \qw      & \ctrl{1} & \targ    & \gate{H}         & \gate{H} & \targ    & \ctrl{3} & \qw      & \qw                   & \qw      & \ctrl{3} & \targ    & \gate{H}         & \qw \\
    \lstick{| q_4 = 0 \rangle}    & \qw      & \gate{H} & \qw      & \targ    & \ctrl{1} & \qw                   & \ctrl{1} & \targ    & \qw      & \gate{H}         & \qw      & \qw      & \qw      & \qw      & \qw                   & \qw      & \qw      & \qw      & \qw              & \qw \\
    \lstick{| q_5 = 0 \rangle}    & \qw      & \gate{R} & \qw      & \qw      & \targ    & \gate{R_z (\theta_1)} & \targ    & \qw      & \qw      & \gate{R^\dagger} & \qw      & \qw      & \qw      & \qw      & \qw                   & \qw      & \qw      & \qw      & \qw              & \qw \\
    \lstick{| q_{10} = 0 \rangle} & \qw      & \qw      & \qw      & \qw      & \qw      & \qw                   & \qw      & \qw      & \qw      & \qw              & \gate{H} & \qw      & \targ    & \ctrl{1} & \qw                   & \ctrl{1} & \targ    & \qw      & \gate{H}         & \qw \\  
    \lstick{| q_{11} = 0 \rangle} & \qw      & \qw      & \qw      & \qw      & \qw      & \qw                   & \qw      & \qw      & \qw      & \qw              & \gate{R} & \qw      & \qw      & \targ    & \gate{R_z (\theta_2)} & \targ    & \qw      & \qw      & \gate{R^\dagger} & \qw \\  
} 
} 
\vspace{0.5cm}
\caption{
Circuit $\mathcal{C}^{\mathrm{LiH}}_1$ for preparation of $| \Psi_{\mathrm{ref}} \rangle$ and operation of $U_1 (\theta_1, \theta_2)$ in eq. (\ref{LiH_ansatz_1}).
Each of the twelve qubits is initially set to $| 0 \rangle$. 
$R \equiv R_x (\pi/2)$ is a gate for rotation around the $x$ axis
and
$R_z (\theta)$ is that around the $z$ axis.\cite{Nielsen_and_Chuang}
The qubits $| q_6 \rangle, | q_7 \rangle, | q_8 \rangle$, and $| q_9 \rangle$ are not shown in the figure since they undergo no operation.
} 
\label{LiH_circuit_ansatz_1} 
\end{figure*}
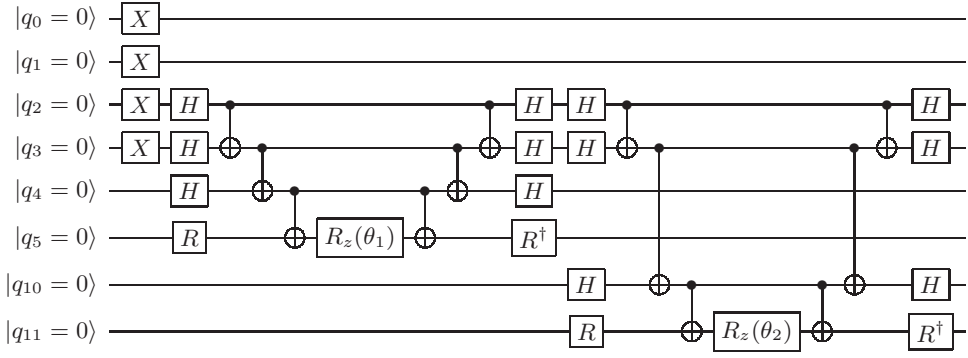

By fixing the bond length at $1.6$ \AA {} in an LiH molecule,
we performed an RHF calculation and obtained $E_{\mathrm{RHF}} = -213.9322$ eV and
six spatial orbitals
among which the two lowest ones were fully occupied.
Therefore we adopted the RHF solution as the reference state
$
| \Psi_{\mathrm{ref}} \rangle
=
a^\dagger_{1 \downarrow}
a^\dagger_{1 \uparrow}
a^\dagger_{0 \downarrow}
a^\dagger_{0 \uparrow}
| \mathrm{vac} \rangle
    \propto
        X_3
        X_2
        X_1
        X_0
        | 0 \cdots 0 \rangle
$
in the JW representation,
where $X_j, Y_j,$ and $Z_j$ for $j = 0, \dots, 11$ are the Pauli matrices acting on the $j$th qubit,
for the subsequent simulations of quantum computation with twelve qubits for the STO-3G basis (twelve) functions.
We tried two excitation operators
$
T_1 (\theta_1, \theta_2)
=
    \theta_1
    a^\dagger_{2 \downarrow}
    a^\dagger_{2 \uparrow}
    a_{1 \downarrow}
    a_{1 \uparrow}
    +
    \theta_2
    a^\dagger_{5 \downarrow}
    a^\dagger_{5 \uparrow}
    a_{1 \downarrow}
    a_{1 \uparrow}
$
and
$
T_2 (\theta_1, \theta_2)
=
    \theta_1
    a^\dagger_{3 \downarrow}
    a^\dagger_{3 \uparrow}
    a_{1 \downarrow}
    a_{1 \uparrow}
    +
    \theta_2
    a^\dagger_{4 \downarrow}
    a^\dagger_{4 \uparrow}
    a_{1 \downarrow}
    a_{1 \uparrow}
,
$
each of which excites the two electrons in the highest occupied molecular orbital (HOMO),
composed mainly of the Li $2 s$ orbital,
to the unoccupied orbital. [See Fig. \ref{fig:orb_energies}(a)]
We rewrite each of $T_1$ and $T_2$ to Pauli tensors for the qubits and pick up only a single tensor from them for each parameter as an approximation similarly to Hempel et al.,\cite{bib:4518}
which is then substituted into eq. (\ref{def_UCC_opr}) to define the ansatz.
The ans\"atze in this case thus read
\begin{gather}
    U_1 (\theta_1, \theta_2)
    \nonumber \\
    =
        \exp
        \left(
            -
            i
            \frac{\theta_2}{2}
            Y_{11}
            X_{10}
            X_{3}
            X_{2}
        \right)
        \exp
        \left(
            -i
            \frac{\theta_1}{2}
            Y_5
            X_4
            X_3
            X_2
        \right)
    \label{LiH_ansatz_1}
\end{gather}
for $T_1$ and
\begin{gather}
    U_2 (\theta_1, \theta_2)
    \nonumber \\
    =
        \exp
        \left(
            -
            i
            \frac{\theta_2}{2}
            Y_{9}
            X_{8}
            X_{3}
            X_{2}
        \right)
        \exp
        \left(
            -i
            \frac{\theta_1}{2}
            Y_7
            X_6
            X_3
            X_2
        \right)
    \label{LiH_ansatz_2}
\end{gather}
for $T_2$,
where we have rescaled the real parameters.
We constructed the circuits $\mathcal{C}^{\mathrm{LiH}}_1$ and $\mathcal{C}^{\mathrm{LiH}}_2$ that act as these unitary operators and optimized the parameters to obtain the UCC ground-state energies.
$\mathcal{C}^{\mathrm{LiH}}_1$
actually operates only on the eight among the twelve qubits,
as shown in Fig. \ref{LiH_circuit_ansatz_1}.
It is similarly the case with $\mathcal{C}^{\mathrm{LiH}}_2$.
The optimized $U_1$ gave $E_{\mathrm{UCC 1}} = -214.3323$ eV,
closer to the FCI value $E_{\mathrm{FCI}} = -214.4889$ eV
than the optimized $U_2$ did with $E_{\mathrm{UCC 2}} = -213.9758$ eV.

\subsubsection{GFs exact within UCC}

\begin{figure}
\begin{center}
\includegraphics[width=6.5cm]{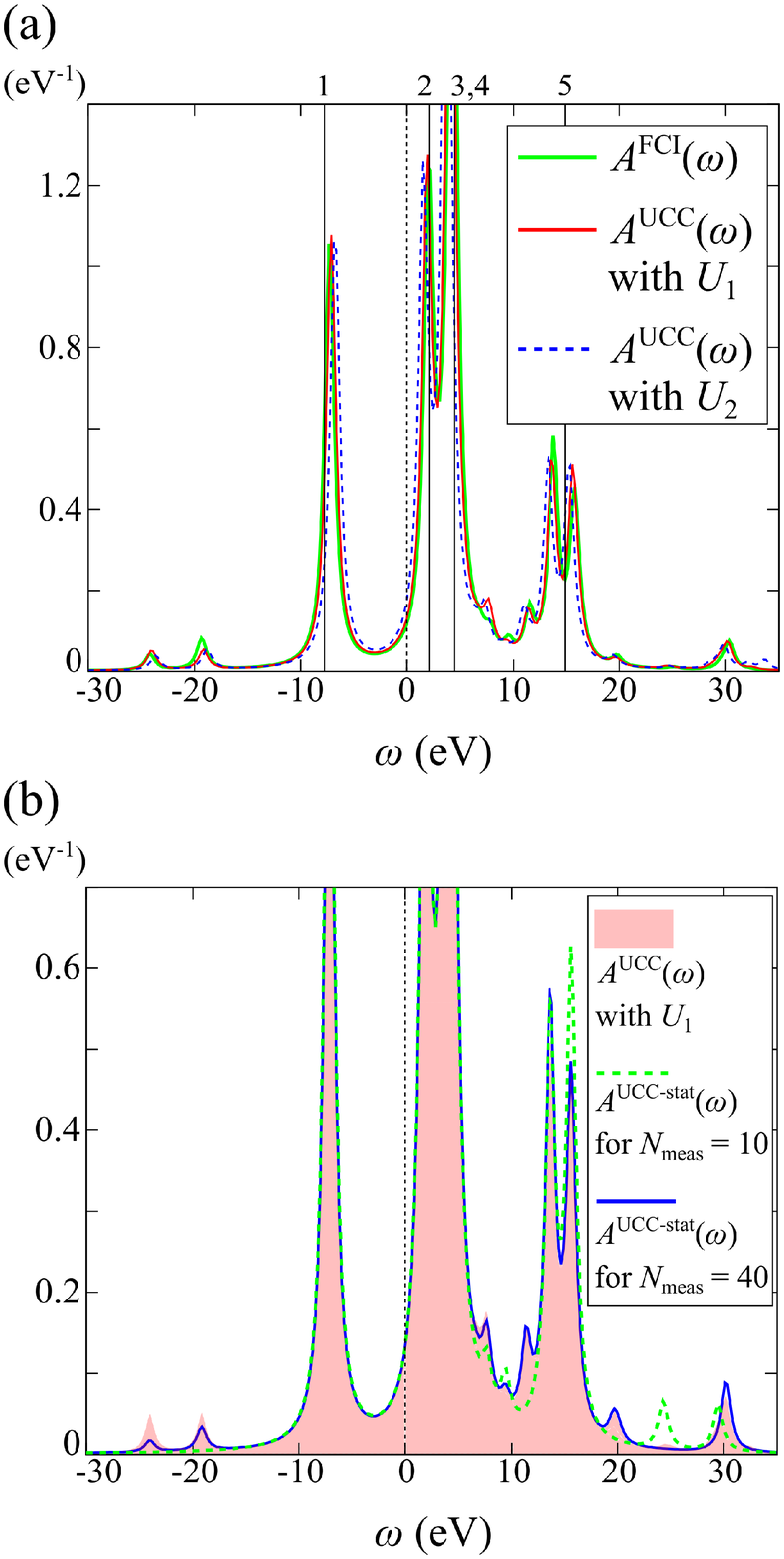}
\end{center}
\caption{
(a)
Spectral functions of an LiH molecule calculated from the ground states of the FCI and optimized UCC solutions.
Solid vertical lines represent the HF orbital energies,
whose indices are also shown near the individual lines.
[See Fig. \ref{fig:orb_energies}(a)]
(b)
For the number of measurements $N_{\mathrm{meas}} = 10$ and $40$ for each component of the GF using the optimized $U_1$,
typical spectral functions $A^{\mathrm{UCC-stat}} (\omega)$ obtained via statistical sampling are shown.
}
\label{fig:LiH_spec}
\end{figure}

We calculated the GFs from the ground states of the FCI and optimized UCC solutions,
as shown in Fig. \ref{fig:LiH_spec} (a).
The FCI spectra $A^{\mathrm{FCI}} (\omega)$ exhibit the weak satellite peaks,
which are correlation effects and thus are absent in the HF spectra.
Specifically,
the weak peaks are seen for $-30 < \omega < -15$ eV, $5 < \omega < 12.5$ eV,
and $17.5$ eV $< \omega$.
Although the major peaks,
called the quasiparticle peaks,
can be basically assigned to the individual HF orbitals,
the two neighboring major peaks around $\omega = 15$ are split due to the correlation effects on the HF orbital $5$.
The satellite peaks are also seen in the UCC spectra $A^{\mathrm{UCC}} (\omega)$ for both $U_1$ and $U_2$.
The quasiparticle peaks in the FCI spectra are closer to the Fermi level ($\omega = 0$) than the HF orbital energies are,
which is due to the well known fact that HF solutions overestimate energy gaps in general.
The overall shapes of the FCI and UCC spectra look quite similar to each other despite the simple ans\"atze
since an LiH molecule is a weakly correlated system.
The locations of quasiparticle and satellite peaks in the UCC spectra for the optimized $U_1$ are closer to those in the FCI spectra than those for the optimized $U_2$,
as expected.

\subsubsection{UCC GFs via statistical sampling}

Hereafter we denote the ground state for the optimized $U_1$ simply by the UCC ground state $| \Psi_{\mathrm{gs}}^{N (\mathrm{UCC})} \rangle$.
To simulate the scheme for obtaining GFs on a quantum computer proposed above,
we calculated the transition matrix elements between $| \Psi_{\mathrm{gs}}^{N (\mathrm{UCC})} \rangle$ and the FCI energy eigenstates $| \Psi_{\lambda}^{N \pm 1 (\mathrm{FCI})} \rangle$. 
We generated random numbers according to these values
since they represent the probability distributions of the measurement results for the qubits.
[See eqs. (\ref{prob_energy_diag_el})-(\ref{prob_energy_off_diag_hole})]
By building the histograms of the results of simulated measurements,
we constructed the GF $G^{\mathrm{UCC-stat}}$ for the UCC ground state.
We denote such construction of all the components of a GF by a single simulation of GF in what follows.

Typical spectral functions $A^{\mathrm{UCC-stat}} (\omega)$ simulated
in this way are shown in Fig. \ref{fig:LiH_spec}(b).
We can see that the quasiparticle peaks in $A^{\mathrm{UCC}} (\omega)$  are well reproduced by the statistical sampling even for the smaller $N_{\mathrm{meas}}$.
For the satellite peaks, on the other hand,
their shapes for the two values of $N_{\mathrm{meas}}$ can be quite different from each other.
In particular, those near $\omega = -25$ and $20$ eV were not even detected for $N_{\mathrm{meas}} = 10$ due to the too few measurements.
These observations indicate that a number of measurements on a quantum computer have to be performed if one wants to capture the correlation effects accurately,
just as PES experiments and their inverse have to be conducted many times for the rare physical processes.

\begin{figure}
\begin{center}
\includegraphics[width=5cm]{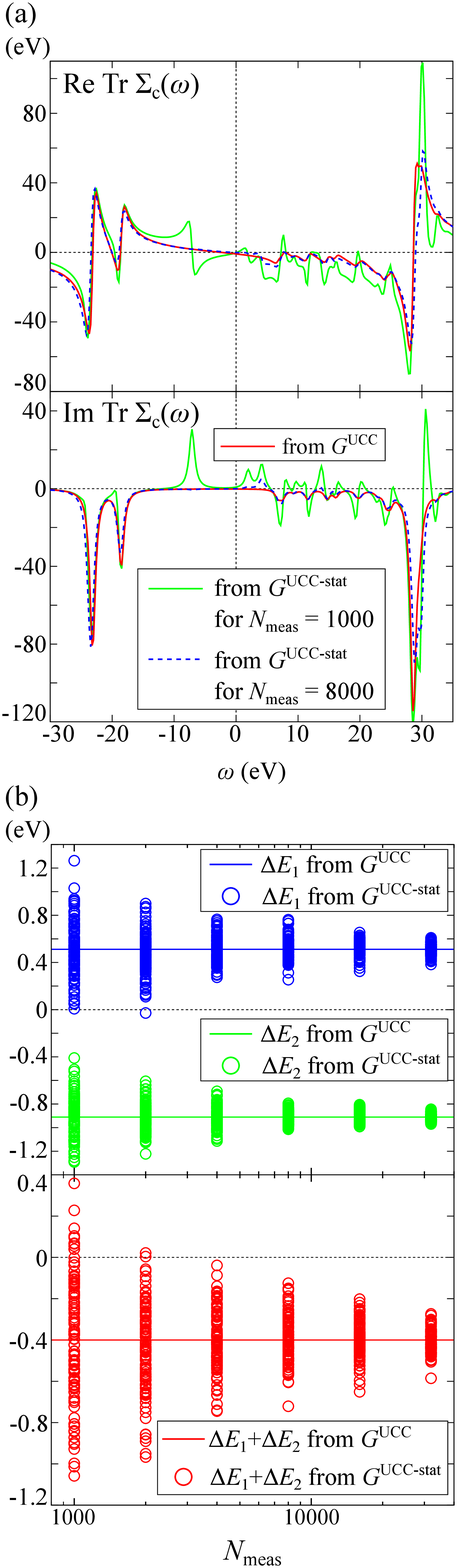}
\end{center}
\caption{
(a)
Typical self-energies of an LiH molecule calculated from sampled UCC GFs, $G^{\mathrm{UCC-stat}}$
,
for $N_{\mathrm{meas}} = 1000$ and $8000$.
The self-energy calculated from $G^{\mathrm{UCC}}$ is also plotted as solid curves.
(b) Correlation energies calculated from $G^{\mathrm{UCC-stat}}$ for the GM formula.
Each circle represents a single simulation in which $N_{\mathrm{meas}}$ measurements were performed for each component of the GF.
Those calculated from $G^{\mathrm{UCC}}$ are also shown as horizontal lines.
}
\label{fig:LiH_self_gm}
\end{figure}

Figure \ref{fig:LiH_self_gm}(a) shows 
the typical shapes of the traces of self-energies $\Sigma_{\mathrm{c}}$ calculated from $G^{\mathrm{UCC-stat}}$
with $N_{\mathrm{meas}} = 1000$ and $8000$.
We notice that the convergence of self-energy with respect to $N_{\mathrm{meas}}$ looks far from satisfaction even for $N_{\mathrm{meas}}= 1000$,
in contrast to the sampled GF. [See Fig. \ref{fig:LiH_spec} (b)]
This observation comes from the fact that
the major contributions to the GF, nothing but the quasiparticle peaks,
are already taken into account as the HF GF,
while the presence of $\Sigma_{\mathrm{c}}$ results solely from the correlation effects.

\subsubsection{Correlation energy from GF}

To examine the statistical behavior of $G^{\mathrm{UCC-stat}}$ quantitatively,
we performed 100 simulations to obtain $G^{\mathrm{UCC-stat}}$ for each given value of $N_{\mathrm{meas}}$ and calculated the correlation energies by using the GM formula in eqs. (\ref{GM_formula_Delta_E_1}) and (\ref{GM_formula_Delta_E_2}).
The results for $N_{\mathrm{meas}} = 1000, 2000, 4000, 8000, 16000$, and $32000$ are shown in Fig. \ref{fig:LiH_self_gm}(b), where
$\Delta E_1 [G^{\mathrm{UCC-stat}}]$
and
$\Delta E_2 [G^{\mathrm{UCC-stat}}]$
scatter around the ideal values,
$\Delta E_1 [G^{\mathrm{UCC}}]$
and
$\Delta E_2 [G^{\mathrm{UCC}}]$, respectively.
The deviations of the sampled values from the ideal values decrease as $N_{\mathrm{meas}}$ increases, as expected.

\subsection{H$_2$O molecule}

\subsubsection{UCC calculations}

By fixing the O-H bond length at $0.96$ \AA {} and the H-O-H bond angle at $104.5^\circ$ in an H$_2$O molecule,
we performed an RHF calculation and obtained $E_{\mathrm{RHF}} = -2039.8504$ eV and
seven spatial orbitals
among which the five lowest ones were fully occupied.
Therefore we adopted the RHF solution as the reference state
$
| \Psi_{\mathrm{ref}} \rangle
\propto
X_9
\cdots
X_0
| 0 \cdots 0 \rangle
$
in the JW representation,
for the subsequent simulations of quantum computation with fourteen qubits for the STO-3G basis (fourteen) functions.
We tried two excitation operators
\begin{gather}
T_2 (\theta_1, \dots, \theta_4)
=
    \theta_1
    a^\dagger_{5 \downarrow}
    a^\dagger_{5 \uparrow}
    a_{3 \downarrow}
    a_{3 \uparrow}
    +
    \theta_2
    a^\dagger_{6 \downarrow}
    a^\dagger_{6 \uparrow}
    a_{3 \downarrow}
    a_{3 \uparrow}
    \nonumber \\
    +
    \theta_3
    a^\dagger_{5 \downarrow}
    a^\dagger_{5 \uparrow}
    a_{4 \downarrow}
    a_{4 \uparrow}
    +
    \theta_4
    a^\dagger_{6 \downarrow}
    a^\dagger_{6 \uparrow}
    a_{4 \downarrow}
    a_{4 \uparrow}
\end{gather}
and
\begin{gather}
T_1 (\theta_1, \dots, \theta_6)
=
    T_2 (\theta_1, \dots, \theta_4)
    \nonumber \\
    +
    \theta_5
    a^\dagger_{5 \downarrow}
    a^\dagger_{5 \uparrow}
    a_{2 \downarrow}
    a_{2 \uparrow}
    +
    \theta_6
    a^\dagger_{6 \downarrow}
    a^\dagger_{6 \uparrow}
    a_{2 \downarrow}
    a_{2 \uparrow}
,
\end{gather}
each of which excites the electrons in the MOs near the Fermi level,
composed mainly of the O $2 p$ orbitals,
to the unoccupied orbitals. [See Fig. \ref{fig:orb_energies}(b)]
We rewrite them to the qubit operators with approximations similarly to the case of an LiH molecule
and introduced the ans\"atze
\begin{gather}
    U_2 (\theta_1, \dots, \theta_4)
    =
        \exp
        \left(
            -i
            \frac{\theta_4}{2}
            Y_{13}
            X_{12}
            X_9
            X_8
        \right)
    \cdot
    \nonumber \\
    \cdot
        \exp
        \left(
            -i
            \frac{\theta_3}{2}
            Y_{11}
            X_{10}
            X_9
            X_8
        \right)
        \exp
        \left(
            -i
            \frac{\theta_2}{2}
            Y_{13}
            X_{12}
            X_7
            X_6
        \right)
    \cdot
    \nonumber \\
    \cdot
        \exp
        \left(
            -i
            \frac{\theta_1}{2}
            Y_{11}
            X_{10}
            X_7
            X_6
        \right)
    \label{H2O_ansatz_2}
\end{gather}
for $T_2$ and
\begin{gather}
    U_1 (\theta_1, \dots, \theta_6)
    =
        \exp
        \left(
            -i
            \frac{\theta_6}{2}
            Y_{13}
            X_{12}
            X_5
            X_4
        \right)
    \cdot
    \nonumber \\
    \cdot
        \exp
        \left(
            -i
            \frac{\theta_5}{2}
            Y_{11}
            X_{10}
            X_5
            X_4
        \right)
        U_2 (\theta_1, \dots, \theta_4)
    \label{H2O_ansatz_1}
\end{gather}
for $T_1$,
where we have rescaled the real parameters.
We constructed the circuits $\mathcal{C}^{\mathrm{H}_2 \mathrm{O}}_1$ and $\mathcal{C}^{\mathrm{H}_2  \mathrm{O}}_2$ that act as these unitary operators and optimized the parameters to obtain the UCC ground-state energies.
The optimized $U_1$ gave $E_{\mathrm{UCC 1}} = -2040.4359$ eV,
closer to the FCI value $E_{\mathrm{FCI}} = -2041.2013$ eV
than the optimized $U_2$ did with $E_{\mathrm{UCC 2}} = -2040.0492$ eV.

\subsubsection{GFs exact within UCC}

\begin{figure}
\begin{center}
\includegraphics[width=6.5cm]{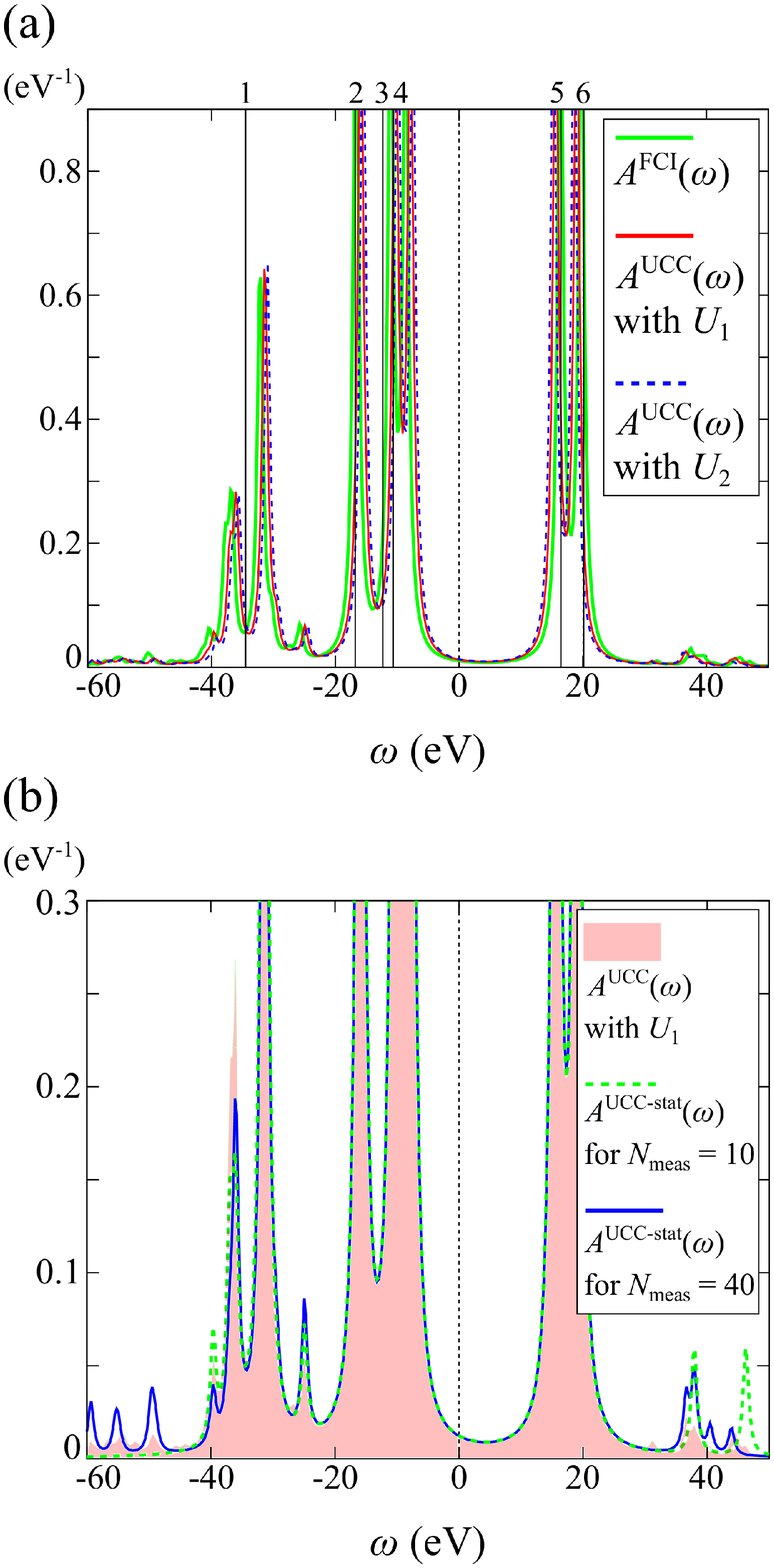}
\end{center}
\caption{
(a)
Spectral functions of an H$_2$O molecule calculated from the ground states of the FCI and optimized UCC solutions.
Solid vertical lines represent the HF orbital energies,
whose indices are also shown near the individual lines. [See Fig. \ref{fig:orb_energies}(b)]
(b)
Typical spectral functions $A^{\mathrm{UCC-stat}} (\omega)$ obtained via statistical sampling,
similarly to Fig. \ref{fig:LiH_spec}(b).
}
\label{fig:H2O_spec}
\end{figure}

We calculated the GFs from the ground states of the FCI and optimized UCC solutions,
as shown in Fig. \ref{fig:H2O_spec} (a).
These three spectral functions admit analyses similar to those for an LiH molecule described above,
since an H$_2$O molecule is also a weakly correlated system.

\subsubsection{UCC GFs via statistical sampling}

Hereafter we denote the ground state for the optimized $U_1$ simply by the UCC ground state.
We performed simulations for obtaining GFs via statistical sampling in the same manner as in the case of an LiH molecule.
Typical simulated spectral functions are shown in Fig. \ref{fig:H2O_spec}(b).
We can see that the quasiparticle peaks in $A^{\mathrm{UCC}} (\omega)$ are well reproduced by the statistical sampling,
while the sampled satellite peaks are not satisfactory.
These results are similar to those in the LiH case.

\begin{figure}
\begin{center}
\includegraphics[width=5cm]{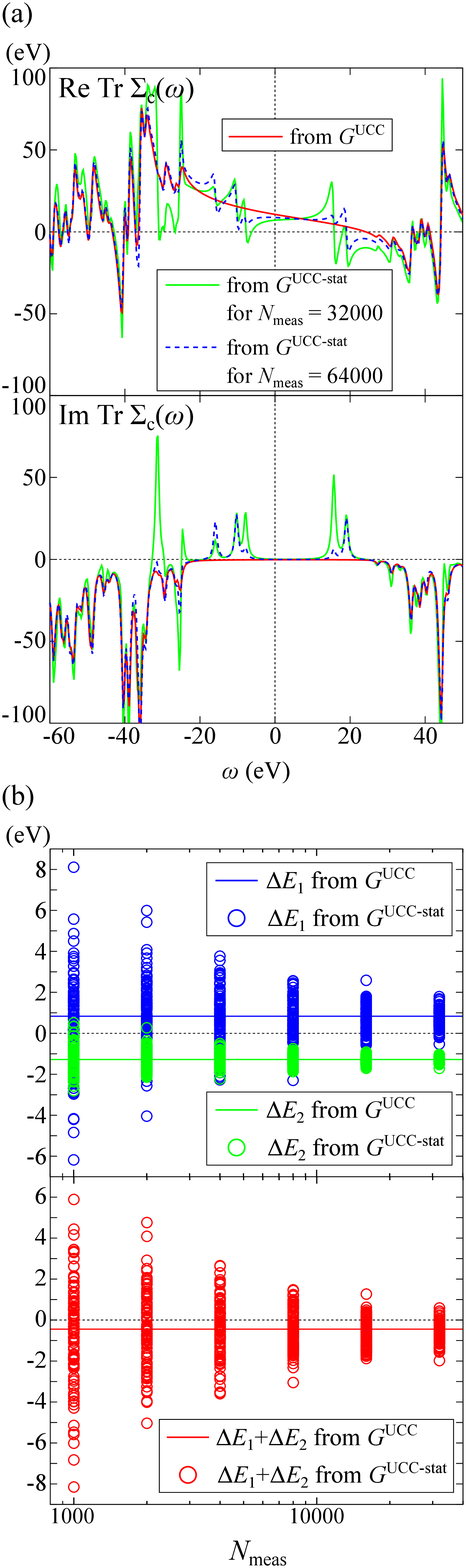}
\end{center}
\caption{
(a)
Typical self-energies of an H$_2$O molecule calculated from sampled UCC GFs, $G^{\mathrm{UCC-stat}}$,
similarly to Fig. \ref{fig:LiH_self_gm}(a).
(b) Correlation energies calculated from $G^{\mathrm{UCC-stat}}$ for the GM formula,
similarly to Fig. \ref{fig:LiH_self_gm}(b).
}
\label{fig:H2O_self_gm}
\end{figure}

Figure \ref{fig:H2O_self_gm}(a) shows 
the typical shapes of the traces of self-energies $\Sigma_{\mathrm{c}}$ calculated from $G^{\mathrm{UCC-stat}}$
with $N_{\mathrm{meas}} = 32000$ and $64000$.
The convergence of self-energy with respect to $N_{\mathrm{meas}}$ is found to be much slower than in the LiH case.
This slow convergence propagates to that of the sampled correlation energies, as explained below.

\subsubsection{Correlation energy from GF}

Similarly to the case of an LiH molecule,
we performed 100 simulations to obtain $G^{\mathrm{UCC-stat}}$ for each given value of $N_{\mathrm{meas}}$ and calculated the correlation energies by using the GM formula,
as shown in Fig. \ref{fig:H2O_self_gm}(b).
Although the increase in $N_{\mathrm{meas}}$ leads to the convergence of sampled correlation energy as well as for an LiH molecule,
the convergence for this case is much slower.
$N_{\mathrm{meas}} = 32000$ achieves the convergence of $\Delta E_1 + \Delta E_2$ within about $0.2$ eV accuracy for an LiH molecule
[see  Fig. \ref{fig:LiH_self_gm}(b)],
while the same $N_{\mathrm{meas}}$ only achieves an accuracy as large as $1.5$ eV for an H$_2$O molecule.
These observations reflect the generic fact that 
the increase in the number of electrons immediately means the rapid increase in the excitation channels,
which forces us to perform measurements on a quantum computer much more times to reproduce the correct probability distribution.
Although the accuracy achieved in our simulations is far from the chemical accuracy, 1 kcal/mol $\approx$ 43 meV,
it seems that we are left with much room for improving the na\"ive scheme proposed in the present study.
In particular,
the pursuit of efficient construction of histograms leading to the suppression of the rapid increase in the necessary number of measurements is valuable in the future.

\section{Conclusions}
\label{sec:conclusions}

We proposed a scheme for the construction of one-particle GF of an interacting electronic system via statistical sampling on a quantum computer.
We were able to circumvent the restriction of unitarity of qubit operations by introducing the quantum circuits for probabilistic state preparation.
The quantum circuits for the diagonal and off-diagonal components and the subsequent QPE allow us to calculate the GF straightforwardly from the histogram obtained via measurements on the qubits.

For demonstrating the validity of our scheme,
we performed simulations of such construction of GFs for LiH and H$_2$O molecules based on the UCC method
by referring to the spectral functions exact within UCC.
We found that the accurate reproduction of weaker satellite peaks requires more measurements to detect the small contributions to the spectra.
We also examined the accuracy of sampling method by exploiting the GM formula to find that
the increase in the number of electrons leads to the rapid increase in the excitation channels,
which forces us to perform measurements many times to 
get a correct histogram.

We should keep in mind that our simulations were performed on the assumption that
the many-electron energy eigenvalues of the target systems are known and
the QPE experiments are conducted with no probabilistic error. 
The results in the present study thus indicate that
the requirements of resources for the accurate description of correlation effects using a real quantum computer grow rapidly as the target systems become large,
as long as we use the simple statistical sampling.
Therefore we have to improve the scheme for obtaining GFs accurately by considering more realistic setups
and simultaneously reducing costs in the future.

\begin{acknowledgments}
This research was supported by MEXT as Exploratory Challenge on Post-K computer (Frontiers of Basic Science: Challenging the Limits) and Grants-in-Aid for Scientific Research (A) (Grant Numbers 18H03770) from JSPS (Japan Society for the Promotion of Science).
\end{acknowledgments}

\begin{widetext}

\appendix

\section{pseudocodes for GF}
\label{appendix:pseudocodes}

Here we provide the pseudocodes for the calculation process of GF proposed in the present study.
We assume that
not only the energy of $N$-electron ground state
but also the energy eigenvalues of $(N \pm 1)$-electron states 
have been obtained before entering the calculation process for GF.
The main process, \textsc{CalcGF}, is given by procedure \ref{alg:CalcGF}.
\textsc{CalcAmplsDiag} in procedure \ref{alg:CalcAmplsDiag} is called to calculate the diagonal components of transition matrices,
while
\textsc{CalcAmplsOffDiag} in procedure \ref{alg:CalcAmplsOffDiag} is called to calculate the off-diagonal components.
The latter calls \textsc{CalcAmplsAux} in procedure \ref{alg:CalcAmplsAux} to get $D^{(\mathrm{e,h}) \pm}_{m m'}$,
from which the off-diagonal components $B^{(\mathrm{e,h})}_{m m'}$ are calculated using eq. (\ref{transition_G_e_off_diag_alpha_half}).

\begin{algorithm}[H]
	\caption{Calculation of GF via statistical sampling}
    \label{alg:CalcGF}
	\begin{algorithmic}[1]
    	\Require 
    	    \Statex {
    	        Hamiltonian $\mathcal{H}$,
    	        number of spatial orbitals $n_{\mathrm{orbs}}$,
    	        $N$-electron ground state $| \Psi^N_{\mathrm{gs}} \rangle$
    	        with its energy eigenvalue $E^N_{\mathrm{gs}}$,
    	        energy eigenvalues $E_\lambda^{N \pm 1}$ of $(N \pm 1)$-electron states,
    	        complex frequency $z$,
    	        number of measurements $N_{\mathrm{meas}}$ for each component
    	        }
		\Ensure
        	\Statex {Electron- $G^{(\mathrm{e})} (z)$ and hole-excitation $G^{(\mathrm{h})} (z)$ parts of GF}
		\Function{CalcGF}{$
            \mathcal{H}, 
            n_{\mathrm{orbs}},
            | \Psi^N_{\mathrm{gs}} \rangle,
		    E^N_{\mathrm{gs}},
		    E^{N + 1},
		    E^{N - 1},
		    z, 
		    N_{\mathrm{meas}}
		    $}
    		\For{$m = 1, \dots, 2 n_{\mathrm{orbs}}$}
            \Comment{Diagonal components}
    		    \State $G^{(\mathrm{e})}_{mm} := 0, G^{(\mathrm{h})}_{mm} := 0$
    		    \State $B^{(\mathrm{e})}_{mm}, B^{(\mathrm{h})}_{mm} :=$
    		    \textsc{CalcAmplsDiag}$(\mathcal{H}, | \Psi^N_{\mathrm{gs}} \rangle, E^{N + 1}, E^{N - 1}, m, N_{\mathrm{meas}})$
        		\For{$\lambda$}
        		    \State $\switch G^{(\mathrm{e})}_{mm} += \frac{B^{(\mathrm{e})}_{\lambda mm}}{z + E_{\mathrm{gs}}^N - E^{N + 1}_\lambda}$
        		\EndFor
        		\For{$\lambda$}
        		    \State $\switch G^{(\mathrm{h})}_{mm} += \frac{B^{(\mathrm{h})}_{\lambda mm}}{z - E_{\mathrm{gs}}^N + E^{N - 1}_\lambda}$
        		\EndFor
    		\EndFor
    		\For{$m = 1, \dots, 2 n_{\mathrm{orbs}}$}
            \Comment{Off-diagonal components}
        		\For{$m' = 1, \dots, m - 1$}
        		    \State $G^{(\mathrm{e})}_{m m'} := 0, G^{(\mathrm{e})}_{m' m} := 0$
        		    \State $G^{(\mathrm{h})}_{m m'} := 0, G^{(\mathrm{h})}_{m' m} := 0$
        		    \State $B^{(\mathrm{e})}_{m m'}, B^{(\mathrm{h})}_{m m'} :=$ 
        		    \textsc{CalcAmplsOffDiag}$(\mathcal{H}, | \Psi^N_{\mathrm{gs}} \rangle, E^{N + 1}, E^{N - 1}, m, m', N_{\mathrm{meas}})$
            		\For{$\lambda$}
            		    \State $\switch G^{(\mathrm{e})}_{m m'} += \frac{B^{(\mathrm{e})}_{\lambda m m'}}{z + E_{\mathrm{gs}}^N - E^{N + 1}_\lambda}, \, G^{(\mathrm{e})}_{m' m} += \frac{B^{(\mathrm{e}) *}_{\lambda m m'}}{z + E_{\mathrm{gs}}^N - E^{N + 1}_\lambda}$
            		\EndFor
            		\For{$\lambda$}
            		    \State $\switch G^{(\mathrm{h})}_{m m'} += \frac{B^{(\mathrm{h})}_{\lambda m m'}}{z - E_{\mathrm{gs}}^N + E^{N - 1}_\lambda}, \, G^{(\mathrm{h})}_{m' m} += \frac{B^{(\mathrm{h}) *}_{\lambda m m'}}{z - E_{\mathrm{gs}}^N + E^{N - 1}_\lambda}$
            		\EndFor
    		    \EndFor
    		\EndFor
            \State \Return $G^{(\mathrm{e})}, G^{(\mathrm{h})}$
        \EndFunction
	\end{algorithmic}
\end{algorithm}

\begin{algorithm}[H]
	\caption{Calculation of diagonal components of transition matrices}
    \label{alg:CalcAmplsDiag}
	\begin{algorithmic}[1]
		\Function{CalcAmplsDiag}{$
		    \mathcal{H},
		    | \Psi^N_{\mathrm{gs}} \rangle,
		    E^{N + 1},
		    E^{N - 1},
		    m,
		    N_{\mathrm{meas}}
		    $}
            \State $B^{(\mathrm{e})}_{mm} := 0, B^{(\mathrm{h})}_{mm} := 0$
    		\For{$i = 1, \dots, N_{\mathrm{meas}}$}
    		    \State Input $| \Psi^N_{\mathrm{gs}} \rangle$ to $\mathcal{C}_m$ and measure the ancilla
    		    \State $| q^{\mathrm{A}} \rangle :=$ observed ancillary state
    		    \State $E :=$ QPE$(| \widetilde{\Psi} \rangle, \mathcal{H})$
    		    \Comment{For the register $| \widetilde{\Psi} \rangle$ coming out of $\mathcal{C}_m$}
    		    \If{$| q^{\mathrm{A}} \rangle == | 0 \rangle$}
    		        \State Find $E$ among $\{ E_\lambda^{N - 1} \}_\lambda$
                    \State $\switch B^{(\mathrm{h})}_{\lambda mm} += 1$
                \Else
    		        \State Find $E$ among $\{ E_\lambda^{N + 1} \}_\lambda$
                    \State $\switch B^{(\mathrm{e})}_{\lambda mm} += 1$
    		    \EndIf
    		\EndFor
            \State $\switch
                B^{(\mathrm{e})}_{mm} *= 1/N_{\mathrm{meas}},
                B^{(\mathrm{h})}_{mm} *= 1/N_{\mathrm{meas}}
                $
            \State \Return $B^{(\mathrm{e})}_{mm}, B^{(\mathrm{h})}_{mm}$
        \EndFunction
	\end{algorithmic}
\end{algorithm}

\begin{algorithm}[H]
	\caption{Calculation of off-diagonal components of transition matrices from $D^{(\mathrm{e,h}) \pm}_{m m'}$}
    \label{alg:CalcAmplsOffDiag}
	\begin{algorithmic}[1]
		\Function{CalcAmplsOffDiag}{$
		    \mathcal{H},
		    | \Psi^N_{\mathrm{gs}} \rangle,
		    E^{N + 1},
		    E^{N - 1},
		    m,
		    m',
		    N_{\mathrm{meas}}
		    $}
		    \State $D^{(\mathrm{e}) \pm}_{m m'}, D^{(\mathrm{h}) \pm}_{m m'} :=$
		    \textsc{CalcAmplsAux}$(\mathcal{H}, | \Psi^N_{\mathrm{gs}} \rangle, E^{N + 1}, E^{N - 1}, m, m', N_{\mathrm{meas}})$
		    \State $D^{(\mathrm{e}) \pm}_{m' m}, D^{(\mathrm{h}) \pm}_{m' m} :=$
		    \textsc{CalcAmplsAux}$(\mathcal{H}, | \Psi^N_{\mathrm{gs}} \rangle, E^{N + 1}, E^{N - 1}, m', m, N_{\mathrm{meas}})$
    		\For{$\lambda$}
                \State $B_{\lambda m m'}^{\mathrm{(e)}}
                    :=
                        e^{-i \pi/4}
                        (
                            D_{\lambda m m'}^{\mathrm{(e)} +}
                            -
                            D_{\lambda m m'}^{\mathrm{(e)} -}
                        )
                        +
                        e^{i \pi/4}
                        (
                            D_{\lambda m' m}^{\mathrm{(e)} +}
                            -
                            D_{\lambda m' m}^{\mathrm{(e)} -}
                        )$    		    
            \EndFor
    		\For{$\lambda$}
                \State $B_{\lambda m m'}^{\mathrm{(h)}}
                    :=
                        e^{-i \pi/4}
                        (
                            D_{\lambda m m'}^{\mathrm{(h)} +}
                            -
                            D_{\lambda m m'}^{\mathrm{(h)} -}
                        )
                        +
                        e^{i \pi/4}
                        (
                            D_{\lambda m' m}^{\mathrm{(h)} +}
                            -
                            D_{\lambda m' m}^{\mathrm{(h)} -}
                        )$    		    
            \EndFor
            \State \Return $B^{(\mathrm{e})}_{m m'}, B^{(\mathrm{h})}_{m m'}$
        \EndFunction
	\end{algorithmic}
\end{algorithm}

\begin{algorithm}[H]
	\caption{Calculation of $D^{(\mathrm{e,h}) \pm}_{m m'}$ for off-diagonal components of transition matrices}
    \label{alg:CalcAmplsAux}
	\begin{algorithmic}[1]
		\Function{CalcAmplsAux}{$
		    \mathcal{H},
		    | \Psi^N_{\mathrm{gs}} \rangle,
		    E^{N + 1},
		    E^{N - 1},
		    m,
		    m',
		    N_{\mathrm{meas}}
		    $}
            \State $D^{(\mathrm{e}) \pm}_{m m'} := 0, D^{(\mathrm{h}) \pm}_{m m'} := 0$
    		\For{$i = 1, \dots, N_{\mathrm{meas}}$}
    		    \State Input $| \Psi^N_{\mathrm{gs}} \rangle$ to $\mathcal{C}_{m m'}$ and measure the ancillae
    		    \State $| q_1^{\mathrm{A}} \rangle \otimes | q_0^{\mathrm{A}} \rangle :=$ observed ancillary state
    		    \State $E :=$ QPE$(| \widetilde{\Psi} \rangle, \mathcal{H})$
    		    \Comment{For the register $| \widetilde{\Psi} \rangle$ coming out of $\mathcal{C}_{m m'}$}
                \If{$| q_0^{\mathrm{A}} \rangle  == | 0 \rangle$}
    		        \State Find $E$ among $\{ E_\lambda^{N - 1} \}_\lambda$
        		    \If{$| q_1^{\mathrm{A}} \rangle  == | 0 \rangle$}
                        \State $\switch D^{(\mathrm{h}) +}_{\lambda m m'} += 1$
                    \Else
                        \State $\switch D^{(\mathrm{h}) -}_{\lambda m m'} += 1$
                    \EndIf
                \Else
    		        \State Find $E$ among $\{ E_\lambda^{N + 1} \}_\lambda$
        		    \If{$| q_1^{\mathrm{A}} \rangle  == | 0 \rangle$}
                        \State $\switch D^{(\mathrm{e}) +}_{\lambda m m'} += 1$
                    \Else
                        \State $\switch D^{(\mathrm{e}) -}_{\lambda m m'} += 1$
                    \EndIf
    		    \EndIf
    		\EndFor
            \State $\switch
                D^{(\mathrm{e}) \pm}_{m m'} *= 1/N_{\mathrm{meas}},
                D^{(\mathrm{h}) \pm}_{m m'} *= 1/N_{\mathrm{meas}}
                $
            \State \Return $D^{(\mathrm{e}) \pm}_{m m'}, D^{(\mathrm{h}) \pm}_{m m'}$
        \EndFunction
	\end{algorithmic}
\end{algorithm}

\end{widetext}

\bibliographystyle{apsrev4-1}
\bibliography{ref}

\end{document}